\documentclass[a4paper,twoside,12pt]{article}
\input{obsjkt.sty}

\newcommand{\Msun}{\ensuremath{\,{\rm M}_\odot}}                  
\newcommand{\Rsun}{\ensuremath{\,{\rm R}_\odot}}                  
\newcommand{\rhosun}{\ensuremath{\,\rho_\odot}}                   
\newcommand{\Teff}{\ensuremath{T_{\rm eff}}}                      
\newcommand{\degr}{\ensuremath{^\circ}}                           
\renewcommand{\kms}{\,km\,s$^{-1}$}                               

\newcommand{\Msunnom}{\hbox{$\mathcal{M}^{\rm N}_\odot$}}
\newcommand{\Rsunnom}{\hbox{$\mathcal{R}^{\rm N}_\odot$}}
\newcommand{\Lsunnom}{\hbox{$\mathcal{L}^{\rm N}_\odot$}}

\OBSheader{Rediscussion of eclipsing binaries: zeta Phe}{J.\ Southworth}{2020 Dec}

\begin{document} 

\OBStitle{Rediscussion of eclipsing binaries. Paper I. \\ The totally-eclipsing B-type system zeta Phoenicis}

\OBSauth{John Southworth}

\OBSinstone{Astrophysics Group, Keele University, Staffordshire, ST5 5BG, UK}

\OBSabstract{$\zeta$\,Phe is a bright binary system containing B6\,V and B8\,V stars. It has deep total and annular eclipses, a slightly eccentric orbit with a period of 1.669\,d, apsidal motion and a third body on a wider orbit. The Transiting Exoplanet Survey Satellite light curve and published radial velocities of this system are analysed to determine masses of $3.91 \pm 0.06$\Msun\ and $2.54 \pm 0.03$\Msun\ and radii of $2.84 \pm 0.02$\Rsun\ and $1.89 \pm 0.01$\Rsun. The resulting distance to the system is in agreement with its trigonometric parallax. The physical properties of the stars, with the exception of the effective temperature of the secondary component, can be matched by the predictions of several sets of theoretical stellar evolutionary models for a solar chemical composition and an age of 70--90\,Myr. A spectroscopic analysis of this system is encouraged for the determination of the photospheric chemical composition of the stars, plus improved measurements of their masses and effective temperatures.}


\section*{Detached eclipsing binary star systems}

Eclipsing binary stars form one of the cornerstones of our understanding of the physics of stars, as they are the primary source of direct measurements of stellar masses and radii. Calculation of these properties can be performed using only observational data, geometry, and orbital mechanics \cite{Russell12apj,Hilditch01book}. Of the many types of eclipsing system, perhaps the most important is the detached eclipsing binaries (hereafter dEBs): systems which have experienced no mass transfer so are representative of normal stars.

The measured properties of dEBs have been used to determine how stars evolve \cite{Russell13obs} and check the predictions of theoretical models of stellar structure and evolution \cite{Pols+97mn,Andersen++90apj,Andersen+91aa,Chen+14mn}. The strength and mass dependence of the phenomenon of convective core overshooting has been calibrated using dEBs \cite{Andersen++90apj,Ribas++00mn,Claret07aa,ClaretTorres16aa,ClaretTorres18apj}, empirical relations of the properties of stars have been fitted to the properties of dEBs \cite{Me09mn,Torres++10aarv,Enoch+10aa}, and they have been utilised to determine the primordial helium abundance and the helium-to-metals enrichment ratio \cite{Metcalfe+96apj,Ribas+00mn}.

dEBs are also excellent distance indicators because it is possible to determine the luminosity of a system from the measured radii and effective temperature (\Teff) values of the component stars. They can be used as direct distance indicators if one trusts bolometric corrections calculated from theoretical model atmospheres, because in this case there is no need to calibrate against nearby examples. An alternative approach is to use empirical surface brightness relations calibrated on nearby stars \cite{BarnesEvans76mn,KruszewskiSemeniuk99aca,Me++05aa,Graczyk+17apj} in which case dEBs can be turned into standard candles. The local distance scale now rests on results from the \textit{Gaia} satellite, but dEBs remain useful for determining the distances to nearby galaxies such as the LMC \cite{Pietrzynski+13nat,Pietrzynski+19nat}, SMC \cite{Hilditch++05mn,North++10aa}, M31 \cite{Vilardell+1paa} and M33 \cite{Bonanos+06apj}. In turn, the cosmological distance scale can be anchored on nearby galaxies using distances from dEBs \cite{Freedman+20apj}.

Another use of dEBs is to probe the interior structure of stars either via apsidal motion \cite{ClaretGimenez93aa,ClaretWillems02aa,ClaretGimenez10aa} and other tidal effects \cite{Mazeh08eas}, or by direct determination of the physical properties of a pulsating star. The types of pulsating star found and studied as components of dEBs include $\delta$\,Scuti \cite{Hambleton+13mn,Maceroni+14aa}, $\gamma$\,Doradus \cite{Debosscher+13aa}, $\beta$\,Cepheid \cite{Me+20mn}, SPB \cite{Clausen96aa}, stochastic pulsators \cite{Tkachenko+12mn,Tkachenko+14mn}, EL\,CVn systems \cite{Maxted+13nat,Maxted+14mn}, red giants with solar-like oscillations \cite{Hekker+10apj,Gaulme+13apj,Themessl+18mn} and stars showing tidally-influenced modes \cite{Bowman+19apj,Me+20mn,Handler+20natas,Kurtz+20mn}.



\section*{Current status of the study of dEBs}

dEBs have historically been difficult to study in detail due to the large amount of observing time needed to obtain good light and radial velocity (RV) curves. Compilations of dEBs with precise mass and radius measurements, published by Popper \cite{Popper67araa,Popper80araa}, Andersen \cite{Andersen91aarv} and Torres {\it et al.} \cite{Torres++10aarv} contained 25, 36, 45 and 94 systems, respectively. Harmanec \cite{Harmanec88baicz}, Malkov {\it et al.} \cite{Malkov+06aa}, Eker {\it et al.} \cite{Eker+14pasa}, and others have published larger catalogues with relaxed qualification criteria, but the most useful dEBs for testing theoretical predictions are those for which mass and radius measurements to 2\% (or preferably 1\%) precision are available \cite{Andersen91aarv}.

The situation is now changing due to a deluge of high-precision light curves that are literally falling out of the sky from space telescopes such as CoRoT \cite{Deleuil+18aa}, {\it Kepler}/K2 \cite{Kirk+16aj}  and now TESS (the Transiting Exoplanet Survey Satellite) \cite{Ricker+15jatis}. These missions were all conceived to facilitate the discovery of transiting planetary systems with transit depths as small as 0.01\%, so find it trivial to both detect and produce high-quality light curves of dEBs with eclipse depths up to 50\% or more. In future the PLATO mission \cite{Rauer+14exa} will push the boundaries to even higher-precision data.

Progress has also been made on spectroscopic methods. Many of the dEBs in the catalogues given above were observed using photographic spectroscopy, and the RVs from these studies remain the best available -- an example would be the subject of the current work. More sophisticated approaches, such as the use of large-format CCD detectors and \'echelle spectroscopy on the hardware side, and cross-correlation, spectral disentangling, Doppler tomography and broadening functions on the software side, have greatly improved the measurement precision readily achievable in RV studies. Spectral disentangling \cite{SimonSturm94aa,Hadrava95aas} has several clear advantages: it requires no template, considers all spectra simultaneously, and produces not only the spectroscopic orbits of the stars but their individual spectra in a form suitable for determination of the \Teff\ values and chemical abundances \cite{PavlovskiHensberge05aa,Pavlovski++18mn}.

Konacki {\it et al.} \cite{Konacki+09apj} measured RVs for double-lined binaries to precisions as good as 2\,m\,s$^{-1}$ using a cross-correlation approach. A recent study of the dEB AI\,Phe \cite{Maxted+20mn} used RVs obtained with this method, the TESS light curve of the system, and a wide variety of analyses performed independently by multiple researchers, to determine the masses and radii of the component stars to precisions of 0.2\%. An accompanying work \cite{Miller++20} obtained the \Teff\ values of the two stars to 0.4\% using the \textit{Gaia} parallax and apparent magnitudes of the system.



\section*{Rediscussion of eclipsing binaries}

The author maintains the DEBCat\footnote{\texttt{https://www.astro.keele.ac.uk/jkt/debcat/}} (Detached Eclipsing Binary Catalogue \cite{Me15debcat}) list of dEBs for which masses and radii have been measured to precisions of 2\% or better, although this is not treated as a harsh cut-off. DEBCat was created in 2005 by updating the list of well-studied dEBs given by Andersen \cite{Andersen91aarv} and has since been maintained by including new systems and new results on existing systems.

A cursory inspection of the contents of DEBCat revealed that many of the entries were based on data and analyses obtained several decades ago, and that space-based high-precision light curves are available for most of the systems listed in the catalogue. We expect that new analyses of these systems would in most cases lead to significantly improved physical properties, particularly radii. We therefore decided to commence a series of studies of suitable systems. Most of the new data will come from TESS, as this has a much greater sky coverage than other sources of high-quality light curves, but other databases will be considered where appropriate. No spectroscopic analyses will be performed, in order to allow many dEBs to be studied within the limited time available. Detailed spectroscopic studies of these objects by other workers are highly encouraged in order that precise masses, \Teff\ values and chemical compositions could be measured for many of these systems.

Another consideration is the continued improvement in measurement of the physical constants and solar quantities used in the study of dEBs. The astronomical unit was redefined to be an exact quantity by the International Astronomical Union (IAU) 2012 Resolution B2, and a set of nominal properties of the Sun were delineated in IAU 2015 Resolution B3 \cite{Prsa+16aj}. Although the effects of this are small, they are no longer negligible by current standards. As an example, the radius of the Sun found by Brown \& Christensen-Dalsgaard \cite{BrownChristensen98apj} is smaller by 0.03\% than the value recommended by the IAU. This motivates the systematic reanalysis of dEBs using the same physical constants.

The title chosen for this series of papers is ``Rediscussion of eclipsing binaries'', for two reasons. Firstly, it fits the scientific aims extremely well. Secondly, it references an illustrious series of works produced by Daniel M.\ Popper, beginning with Z\,Her \cite{Popper56apj} and culminating with V380\,Cyg, VV\,Ori and V1765\,Cyg \cite{Popper93pasp}. Our primary aim is the curation of the DEBCat catalogue, but other systems will be covered when interesting results are obtained

In this first work, we consider the bright and early-type dEB $\zeta$\,Phoenicis. This came to our attention whilst searching for a nice totally-eclipsing dEB for an exam question. The TESS light curve is stunning and suitable for a significant improvement in our understanding of the system. It was also found that the uncertainties in the masses had been underestimated in previous works, something that has been noticed for other systems studied using similar spectroscopic material \cite{Gallenne+16aa}.


\section*{The dEB $\zeta$\,Phoenicis}

\begin{table}[t]
\caption{\em Basic information on $\zeta$\,Phe. \label{tab:info}}
\centering
\begin{tabular}{lll}
{\em Property}                      & {\em Value}           & {\em Reference}                   \\[3pt]
Bright Star Catalogue designation   & HR 338                & \cite{HoffleitJaschek91}          \\
Henry Draper designation            & HD 6882               & \cite{CannonPickering18anhar2}    \\
\textit{Hipparcos} designation      & HIP 5348              & \cite{Hip97}                      \\
\textit{Hipparcos} parallax         & $10.92 \pm 0.39$ mas  & \cite{Vanleeuwen07aa}             \\
\textit{Gaia} DR2 ID                & 4913847589156808960   & \cite{Gaia18aa}                   \\
\textit{Gaia} parallax              & $14.68 \pm 0.73$ mas  & \cite{Gaia18aa}                   \\
$B$ magnitude                       & $3.908 \pm 0.014$     & \cite{Hog+00aa}                   \\
$V$ magnitude                       & $4.014 \pm 0.009$     & \cite{Hog+00aa}                   \\
Spectral type                       & B6\,V + B8\,V         & \cite{Popper98pasp}               \\[10pt]
\end{tabular}
\end{table}

$\zeta$\,Phe (Table\,\ref{tab:info}) was reported to be a spectroscopic binary by Wilson \cite{Hogg51mn} from photographic plates taken at Lick Observatory. RVs and a double-lined spectroscopic orbit was presented by Colacevich\cite{Colacevich35pasp} based mostly on the Lick material; \textit{Simbad} misreports the subject of the latter publication as $\delta$\,Phe. Hagemann\cite{Hagemann59mn} obtained 80 photographic spectra and was able to measure RVs for the primary on 71 plates, and for the secondary on only 16 plates. In his \textit{Rediscussion of Eclipsing Binaries, paper 9}, Popper\cite{Popper70apj} presented a spectroscopic orbit of both components of $\zeta$\,Phe based on 21 photographic spectra, finding velocity amplitudes and thus masses significantly different than previous studies. The most recent spectroscopic study of this system \cite{Andersen83aa} was based on 31 high-dispersion spectra obtained using photographic plates, and yielded precise velocity amplitudes for the two stars.

The discovery of eclipses in $\zeta$\,Phe was made by Hogg\cite{Hogg51mn}, who obtained unfiltered photoelectric observations of the star. The primary eclipse is annular and the secondary eclipse is total. Dachs\cite{Dachs71aa} obtained photoelectric observations on the $UBV$ system and Knipe\cite{Knipe71mnssa} observed one primary eclipse. Extensive photometry in the Str\"omgren $uvby$ system was obtained and analysed by the Copenhagen group \cite{Clausen++76aas,Clausen++76aa}. From these results and his own RV analysis, Andersen \cite{Andersen83aa} measured the masses and radii of the two components of the eclipsing system to precisions of approximately 2\%. Since then, a complete light curve has been published by Shobbrook\cite{Shobbrook04jad}.

$\zeta$\,Phe is know to have two fainter nearby companions \cite{Knipe68roci,Vandenbos68roci,Clausen++76aa}: 
a star of magnitude $V=8.2$ at 6.4$^{\prime\prime}$ forms RMK\,2 and one of magnitude $V=6.8$ at 0.5$^{\prime\prime}$ is designated RST\,1205.
A subsequent measurement \cite{Tokovinin+10pasp} gives a separation of 0.5486$^{\prime\prime}$ and a magnitude difference of 2.7\,mag for RST\,1205. Andersen \cite{Andersen83aa} identified RST\,1205 in his spectra, making $\zeta$\,Phe a triple-lined system, and measured its RV from 26 of the plates. The RV of this object was found to be roughly in agreement with the systemic velocity of the eclipsing system, suggesting the three stars are gravitationally bound. This has been confirmed \cite{ZascheWolf07an} with the measurement of the orbital period of the third body ($220.9 \pm 3.5$\,yr) from astrometric observations.

One remaining characteristic of $\zeta$\,Phe to be discussed is the presence of apsidal motion. This was first noticed by Dachs\cite{Dachs71aa}, and an apsidal period of $U = 44.2 \pm 6.5$\,yr was measured by Gim\'enez \textit{et al.} \cite{Gimenez++86aa}. The most recent measurement of the rate of apsidal motion is $\dot\omega = 6.16 \pm 0.20$\,deg\,yr$^{-1}$, corresponding to $U = 58.4 \pm 1.8$\,yr \cite{ZascheWolf07an}.


\section*{Observational material}

\begin{figure}[t] \centering \includegraphics[width=\textwidth]{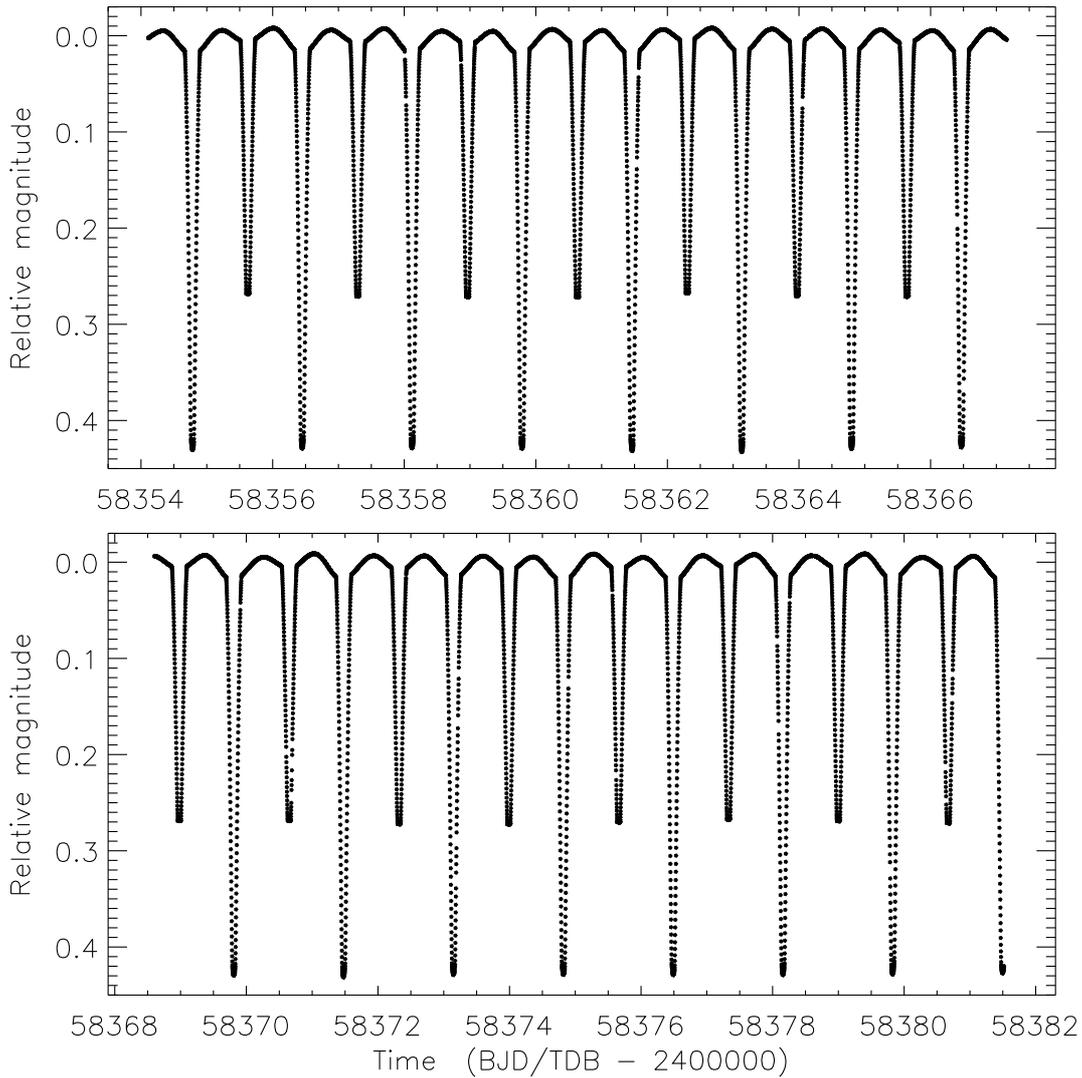} \\
\caption{\label{fig:time} TESS simple aperture photometry of $\zeta$\,Phe from
Sector 2. The upper and lower plots show the observations either side of the
mid-sector pause for data download.} \end{figure}

TESS was launched by NASA on 2018/04/08 into a highly eccentric orbit around the Earth with an orbital period half that of the Moon. It is currently engaged in a photometric survey of 85\% of the celestial sphere with the aim of identifying extrasolar planets through the transit method \cite{Ricker+15jatis}. It includes four cameras with 10.5\,cm apertures, each with four CCDs, that together image a 24$^\circ$$\times$96$^\circ$ strip of sky. Each pixel subtends a solid angle of 21$^{\prime\prime}$$\times$21$^{\prime\prime}$. The observations are performed through wide-band filters with a high response function between 600\,nm and 1000\,nm.

TESS observes individual strips of sky for two orbits (27.4\,d) with a break near the midpoint for download of data to Earth via the NASA Deep Space Network. Such a unit of observation is called a sector, and on its completion TESS moves onto the next sector. As it is designed to detect shallow planet transits, the light curves are of very high quality and thus well suited to studies of large-amplitude variable stars such as dEBs. A total of 200\,000 stars were pre-selected for high-cadence observations (summed into a 120\,s sampling rate). Full-frame images are also captured at a cadence of 1800\,s, and subsequently clipped into an effective integration time of 1425\,s by a cosmic-ray rejection algorithm.

The data are processed and released as light curves by the TESS Science Processing Operations Center for stars observed at high cadence \cite{Jenkins+16spie}. Two versions are available: simple aperture photometry (SAP) and pre-search data conditioning (PDC). The latter is based on the former but undergoes additional processing to remove signals which might obscure shallow transits, a procedure that risks removing astrophysical signal so is not suitable for use on objects showing deep eclipses.

$\zeta$\,Phe was observed\footnote{\texttt{https://heasarc.gsfc.nasa.gov/cgi-bin/tess/webtess/wtv.py}} using TESS in Sector 2 (2018/08/22 to 2018/09/20), in high cadence. These data were downloaded from the Mikulski Archive for Space Telescopes (MAST) archive\footnote{\texttt{https://mast.stsci.edu/portal/Mashup/Clients/Mast/Portal.html}} and the SAP measurements were converted into magnitude units ready for further analysis. Only data with no flagged problems (QUALITY $=$ 0) were retained, and the data uncertainties were ignored because they have little variance and are far too small. Five datapoints were rejected as being 4$\sigma$ outliers, leaving a total of 18\,278 for analysis (Fig.\,\ref{fig:time}).


\section*{Light curve analysis with {\sc jktebop}}

A preliminary analysis of the TESS light curve was performed using version 40 of the {\sc jktebop} code\footnote{\texttt{http://www.astro.keele.ac.uk/jkt/codes/jktebop.html}} \cite{Me08mn,Me13aa}. {\sc jktebop} approximates the stars as spheres for the purposes of calculating eclipse shapes and as ellipsoids for calculation of proximity effects.

We designate the primary star (the one eclipsed during the deeper eclipse) as star A, and the secondary star as star B. In the current case, star A has a larger mass, radius and \Teff\ than star B.

Parameters of the fit included the sum and ratio of the fractional radii of the stars ($r_{\rm A} = \frac{R_{\rm A}}{a}$ and $r_{\rm B} = \frac{R_{\rm B}}{a}$ where $R_{\rm A}$ and $R_{\rm B}$ are the true radii and $a$ is the orbital semimajor axis), the orbital inclination, and the central surface brightness ratio. The orbital period and a reference time of mid-eclipse were also fitted, and no constraints from historical data were applied in order to avoid the complications arising from the known apsidal motion and third-body effects in the system. We included limb darkening using the quadratic law with the linear coefficients fitted and the quadratic coefficients fixed to values from Claret \cite{Claret17aa}. The Poincar\'e elements, $e\cos\omega$ and $e\sin\omega$ where $e$ is the orbital eccentricity and $\omega$ is the argument of periastron, were included as fitted parameters. Finally, third light was included as a fitted parameter because the two fainter nearby stars are much closer than the pixel scale of TESS.

Uncertainties were calculated using Monte Carlo and residual-permutation algorithms \cite{Me08mn}. The results of this process were very encouraging, with uncertainties in $r_{\rm A}$ and $r_{\rm B}$ of roughly 0.1\%. However, $r_{\rm A}$ was found to be sufficiently large that the spherical approximation used in {\sc jktebop} is inaccurate at the level of approximately 1\% \cite{Me+20mn}. This caveat applies also to previous studies that relied on the {\sc ebop} code \cite{Clausen++76aa}. It was therefore necessary to switch to a more sophisticated code.

In preparation for a more refined analysis with a slower code, we used the best-fitting orbital ephemeris from the {\sc jktebop} fit to phase-bin the light curve into 388 datapoints. These were sampled ten times more finely through the eclipses in order to retain all important temporal information in the light curve whilst avoiding the computational expense of calculating a fine grid of out-of-eclipse datapoints.


\section*{Light curve analysis with the Wilson-Devinney code}

\begin{table} \centering
\caption{\em Summary of the parameters for the WD solution of the TESS light curve
of $\zeta$\,Phe. Uncertainties are only quoted when they have been robustly assessed
by comparison between a full set of alternative solutions. \label{tab:wd}}
\begin{tabular}{lcc}
{\em Parameter}                                      & {\em Star A}          & {\em Star B}          \\[3pt]           
{\it Control parameters:} \\                                                                                           %
{\sc wd2004} operation mode                          & \multicolumn{2}{c}{0}                         \\                
Treatment of reflection                              & \multicolumn{2}{c}{1}                         \\                
Number of reflections                                & \multicolumn{2}{c}{1}                         \\                
Limb darkening law                                   & \multicolumn{2}{c}{2 (logarithmic)}           \\                
Numerical grid size (normal)                         & \multicolumn{2}{c}{60}                        \\                
Numerical grid size (coarse)                         & \multicolumn{2}{c}{50}                        \\[3pt]           
{\it Fixed parameters:} \\                                                                                             %
Orbital period (d)                                   & \multicolumn{2}{c}{1.6697739}                 \\                
Primary eclipse time (BJD/TDB)                       & \multicolumn{2}{c}{2458366.46953}             \\                
Mass ratio                                           & \multicolumn{2}{c}{0.649}                     \\                
Rotation rates                                       & 1.0                   & 1.0                   \\                
Gravity darkening                                    & 1.0                   & 1.0                   \\                
\Teff\ (K)                                           & 14\,400               & 12\,000               \\                
Bolometric linear limb darkening coefficient         & 0.7497                & 0.7274                \\                
Bolometric logarithmic limb darkening coeff.         & 0.0709                & 0.0721                \\                
Linear limb darkening coefficient                    & $0.38 \pm 0.17$       & $0.39 \pm 0.10$       \\                
Logarithmic limb darkening coefficient               & 0.1732                & 0.1890                \\[3pt]           
{\it Fitted parameters:} \\                                                                                            %
Phase shift                                          & \multicolumn{2}{c}{$-0.00226 \pm 0.00001$}    \\                
Potential                                            & $4.583 \pm 0.022$     & $4.999 \pm 0.016$     \\                
Orbital inclination (\degr)                          & \multicolumn{2}{c}{$89.14 \pm 0.11$}          \\                
Orbital eccentricity                                 & \multicolumn{2}{c}{$0.0116 \pm 0.0024$}       \\                
Argument of periastron (\degr)                       & \multicolumn{2}{c}{$307 \pm 12$}              \\                
Third light                                          & \multicolumn{2}{c}{$0.1220 \pm 0.0031$}       \\                
Light contributions                                  & $8.464 \pm 0.042$     & $2.573 \pm 0.041$     \\                
Bolometric albedos                                   & 1.0 (fixed)           & 1.47 (fitted)         \\[3pt]           
{\it Derived parameters:} \\                                                                                           %
Light ratio                                          & \multicolumn{2}{c}{$0.3040 \pm 0.0051$}       \\
Fractional radii                                     & $0.2572 \pm 0.0013$   & $0.1710 \pm 0.0006$   \\[10pt]          
\end{tabular}
\end{table}

The phase-binned light curve from the previous section was modelled using the Wilson-Devinney code \cite{WilsonDevinney71apj,Wilson79apj}, which implements Roche geometry to accurately represent the tidally-distorted components of binary star systems. We used the 2004 version of the code \cite{WilsonVanhamme04}, hereafter called {\sc wd2004}, driven via the {\sc jktwd} wrapper \cite{Me+11mn}. Because {\sc wd2004} does not include the TESS passband, it was operated in mode 0 with the \Teff\ values of the stars decoupled from their light contributions.

For the initial fits the fitted parameters were the potentials of the two stars, their light contributions, the amount of third light in the TESS passband, $e$, $\omega$, and the orbital inclination. The rotational velocities of the stars were set to be pseudosynchronous with the orbital motion, the gravity darkening exponents were set to 1.0 \cite{Claret98aas} and the \Teff\ values were fixed at the values given by Andersen \cite{Andersen83aa}. The mass ratio, defined as the ratio of the mass of star B to that of star A, was set to 0.649.

Limb darkening coefficients were taken from Van Hamme \cite{Vanhamme93aj}, and limb darkening was initially specified using the linear law. We immediately found that at least one coefficient for each star had to be varied in order to get a good fit. For reference, the initial values were $u_{\rm A} = 0.196$ and $u_{\rm B} = 0.227$ and the fitted values were $u_{\rm A} = 0.274$ and $u_{\rm B} = 0.464$, so there was a need for stronger limb darkening than theoretically predicted. Further experimentation showed that the square-root limb darkening law was significantly better and that the logarithmic was slightly better still, in all cases with at least one of the two coefficients for each star fitted. For the final model the logarithmic law was adopted and the linear coefficient for each star was fitted. Fitting for both coefficients for each star was not an option as this is not implemented in {\sc wd2004}; it would be very unlikely to help due to the known strong correlations between the coefficients \cite{Me++07aa,Me08mn,Pal08mn,Carter+08apj,Howarth11mn}.

Our attempts to fit for the rotation rate of the primary star yielded values close to pseudosynchronous, and for the secondary star the fits were unstable. In both cases these parameters had no significant effect on the best-fitting fractional radii, which are the two parameters of greatest interest obtainable from the light curve. Fitting for the gravity darkening exponents was similarly unsuccessful.

Fitting for albedo returned a much better fit but for physically unexpected values of the albedos. The best compromise was to fit for the albedo of the secondary component, which always drifted up to values of roughly 1.4, but to fix the albedo for the primary component to 1.0 to avoid getting negative values. Albedo is the ratio of the light emitted to the light incident on the surface of the star, so is expected to be between 0.0 and 1.0 \cite{Claret01mn}. An albedo above unity is therefore not expected, but is physically possible if some flux from shorter wavelengths is reprocessed and emitted in the TESS passband. From the author's experience \cite{Me+11mn,Lehmann+13aa,Me+20mn} albedo can compensate for unrelated inaccuracies of the binary model and often rises above 1.0 when that model is confronted with high-quality data. The treatment of albedo is often a significant contributor to the uncertainty of the final results, and this is the case in the current work.

\begin{figure}[t] \centering \includegraphics[width=\textwidth]{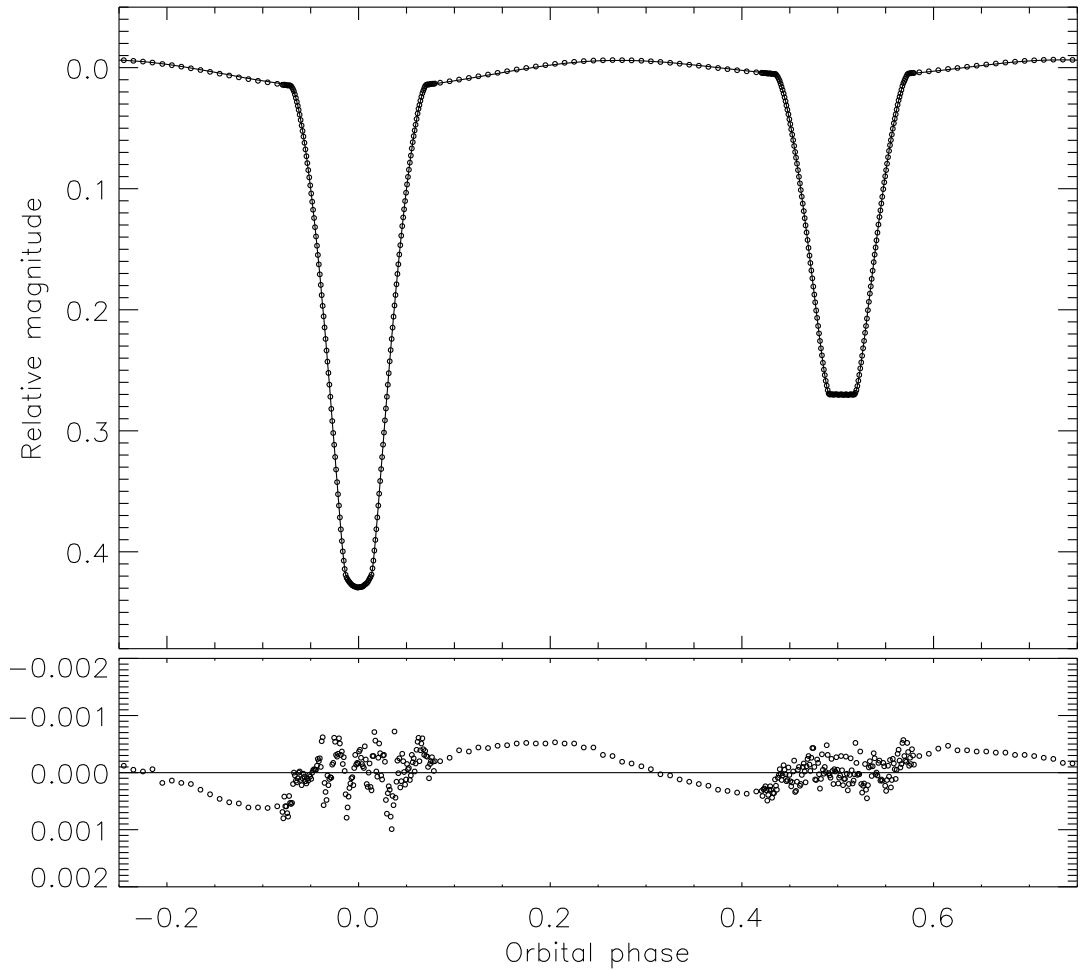} \\
\caption{\label{fig:phase} Best fit to the TESS light curve of $\zeta$\,Phe using {\sc wd2004}.
The phase-binned data are shown using open circles and the best fit with a continuous line.
The residuals are shown on an enlarged scale in the lower panel.} \end{figure}

Table\,\ref{tab:wd} summarises the results of the work above. We designate the final fit to be that with the maximum numerical precision and with logarithmic limb darkening law. The fitted parameters were the potentials, light contributions and linear limb darkening coefficients of the two stars, the orbital eccentricity, argument of periastron, orbital inclination, the phase of primary eclipse, third light, and the albedo of the secondary star. Third light is given as the fraction of the total light of the system at phase 0.25, so is not on the same scale as the light contributions of the two eclipsing stars.

The best fit (Fig.\,\ref{fig:phase}) is good, with an rms versus the phase-binned data of 0.3\,mmag, but is still dominated by systematics. The high-frequency systematics during eclipse can be attributed to the numerical precision of the code. The low-frequency systematics outside eclipse indicate that the proximity effects implemented in {\sc wd2004} are only an approximation and exceed the Poisson noise in this case. In order to test if this could be caused by the Doppler boosting effect \cite{Maxted++00mn,LoebGaudi03apj,Zucker++07apj} the input data were adjusted to account for the expected amount of Doppler boosting using the approach outlined by Loeb \& Gaudi \cite{LoebGaudi03apj} and the fit was rerun. This yielded a slightly worse fit with parameters well within the errorbars given in Table\,\ref{tab:wd}, so this effect is not to blame for the imperfect fit. For the record, the expected amplitude of the Doppler boosting effect for $\zeta$\,Phe is approximately 0.74\,mmag.


\section*{Error analysis for the light curve}

As the Poisson noise in the TESS data is negligible, we have calculated the uncertainties in the fitted parameters by comparing fits obtained with different sets of input parameters. These sets differed from the final fit by changes to the limb darkening law used (linear or square-root versus logarithmic), the way albedo was fitted (primary only, secondary only, both stars), the numerical precision specified in {\sc wd2004}, rotation and gravity darkening. The effect of each of these was determined for each fitted parameter by differencing the fitted value with that from the overall best fit, and these were added in quadrature to give the final uncertainties.

These uncertainties are included in Table\,\ref{tab:wd} and show that the solution is very well-determined. The dominant uncertainty in the fractional radii arises from the treatment of albedo. Until a clear improvement in the understanding of this parameter is obtained, the uncertainties in the fractional radii cannot be significantly lowered. The current uncertainties are only 0.5\% and 0.4\%, so are well within our target of 1\% precision.

{\sc wd2004} computes formal uncertainties from the covariance matrix, and the user guide \cite{WilsonVanhamme04} cautions against their adoption as the true uncertainties of the fitted parameters. The formal uncertainties have been found to be too small in several cases \cite{MaceroniRucinski97pasp,Pavlovski+09mn} and the current analysis allows this to be quantified in the case of negligible Poisson noise in the observational data. The formal uncertainties exceed the true uncertainties for the fitted parameters listed in Table\,\ref{tab:wd} by factors ranging fom 6.6 (for the light contribution of the primary star) to 131 (for the linear limb darkening coefficient of that star). It is clear that the choice of model is a critical contributor to the uncertainties in the solution in data of high quality.

Finally, it is interesting to consider the difference in results between the preliminary fit with {\sc jktebop} and the final fit with {\sc wd2004}. This is restricted to the fractional radii for brevity, for which it amounts to 0.45\% for $r_{\rm A}$ and $-$0.02\% for $r_{\rm B}$. The values for the more distorted primary star differ by roughly the size of the final errorbar, and for the secondary star by a negligible amount. This supports the use of {\sc jktebop} for stars which are relatively undistorted (see also Maxted \textit{et al.} \cite{Maxted+20mn}).


\section*{Physical properties of $\zeta$\,Phe}

\begin{figure}[t] \centering \includegraphics[width=\textwidth]{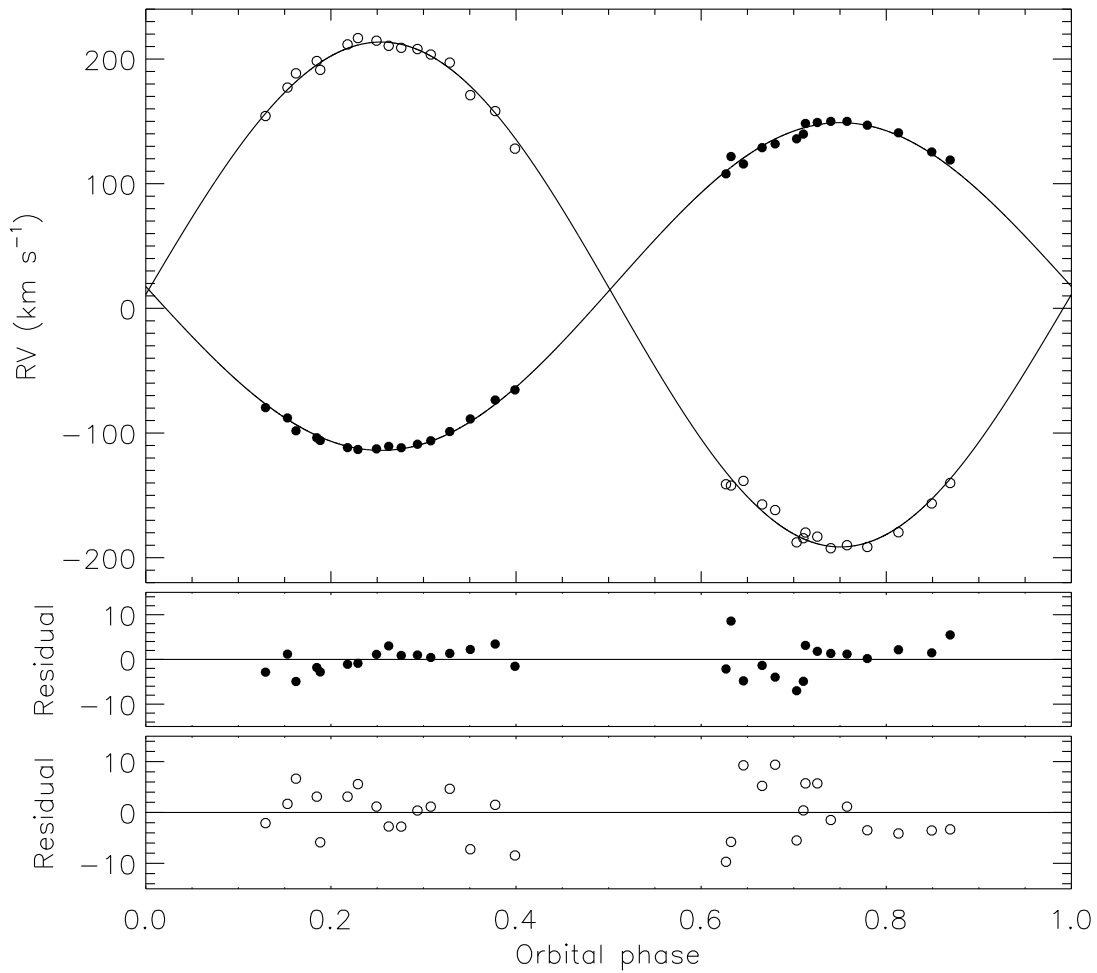} \\
\caption{\label{fig:rvs} Spectroscopic orbit of $\zeta$\,Phe. RVs of the primary and
secondary stars are shown with filled and open circles, respectively. The fitted orbits
are shown using solid lines. The lower panels show the residuals of the fit.} \end{figure}

Armed with the orbital inclination and fractional radii from the light curve analysis above, it is now possible to calculate the full physical properties of the system. This in turn requires some results from spectroscopic analysis: the velocity amplitudes and \Teff\ values of the stars. For the latter we adopted the values proposed by Andersen \cite{Andersen83aa}: $T_{\rm eff,A} = 14\,400 \pm 800$\,K and $T_{\rm eff,B} = 12\,000 \pm 600$\,K.

Velocity amplitudes were also given by Andersen \cite{Andersen83aa}, but these were reanalysed in order to check the uncertainties. The RVs were copied from Andersen \cite{Andersen83aa} and fitted with a Keplerian orbit using {\sc jktebop}. The two stars were not required to have the same systemic velocity, and the fitted values differ by $4.0 \pm 2.1$\kms. The RVs were not supplied with data errors so a single value was chosen for each star to force a reduced $\chi^2$ of $\chi^2_\nu = 1.0$. Thus it was assumed that all RVs for each star are of equal precision, and that the precision can be estimated from the scatter around the best fit. Uncertainties in the velocity amplitudes were calculated using Monte Carlo simulations. The resulting velocity amplitudes are $K_{\rm A} = 131.4 \pm 0.7$\kms\ and $K_{\rm B} = 202.5 \pm 1.3$\kms. The uncertainties in these values are modestly larger than those given by Andersen \cite{Andersen83aa} (0.6 and 1.0\kms, respectively). The RVs and fitted orbits are shown in Fig.\,\ref{fig:rvs}.

The physical properties of the $\zeta$\,Phe system were calculated from the photometric and spectroscopic results using the {\sc jktabsdim} code \cite{Me++05aa}. {\sc jktabsdim} was modified for this work to use the standard physical constants and stellar properties adopted by the International Astronomical Union (IAU) 2012 Resolution B2 and 2015 Resolution B3 \cite{Prsa+16aj}. The uncertainties of all input parameters were propagated to all output parameters using a perturbation analysis. The results are given in Table\,\ref{tab:absdim} and agree well with those from Andersen \cite{Andersen83aa}.

\begin{table} \centering
\caption{\em Physical properties of $\zeta$\,Phe. The \Teff\ values are from Andersen \cite{Andersen83aa}. \label{tab:absdim}}
\begin{tabular}{lr@{\,$\pm$\,}lr@{\,$\pm$\,}l}
{\em Parameter}                           & \multicolumn{2}{c}{\em Star A} & \multicolumn{2}{c}{\em Star B} \\[3pt]
Mass ratio                                & \multicolumn{4}{c}{$0.6490 \pm 0.0053$}       \\
Semimajor axis (\Rsunnom)                 & \multicolumn{4}{c}{$11.022 \pm 0.048$}        \\
Mass (\Msunnom)                           &   3.908 & 0.057       &   2.536 & 0.031       \\
Radius (\Rsunnom)                         &   2.835 & 0.019       &   1.885 & 0.011       \\
Surface gravity ($\log$[cgs])             &  4.1249 & 0.0052      &  4.2917 & 0.0038      \\
Density ($\!$\rhosun)                     &  0.1715 & 0.0027      &  0.3788 & 0.0044      \\
Synchronous rotational velocity (\kms)    &   85.89 & 0.57        &   57.11 & 0.32        \\
\Teff\ (K)                                & 14\,400 & 800         & 12\,000 & 600         \\
Luminosity $\log(L/\Lsunnom)$             &    2.49 & 0.10        &    1.82 & 0.09        \\
$M_{\rm bol}$ (mag)                       & $-$1.49 & 0.24        &    0.19 & 0.21        \\
\end{tabular}
\end{table}

The distance to $\zeta$\,Phe has been measured to be $91.6 \pm 3.2$\,pc, from its trigonometric parallax found using the \textit{Hipparcos} satellite \cite{Vanleeuwen07aa}. The analogous measurement from the \textit{Gaia} satellite \cite{Gaia18aa} is less good because the binary system is so bright. Using the \textit{Hipparcos} $B$ and $V$ magnitudes of the system \cite{Hog+00aa} and bolometric corrections from Girardi \textit{et al.} \cite{Girardi+02aa}, we find a distance to $\zeta$\,Phe of $87.3 \pm 3.3$\,pc. This consistency check is sufficiently successful to support the \Teff\ values used for the stars; an increase of 800\,K is needed to bring the two distances into exact agreement. We have not been able to find reliable photometry in other widely-used passbands, limiting the distance comparisons that can be made.


\section*{Comparison with theoretical models}

\begin{figure}[t] \centering \includegraphics[width=\textwidth]{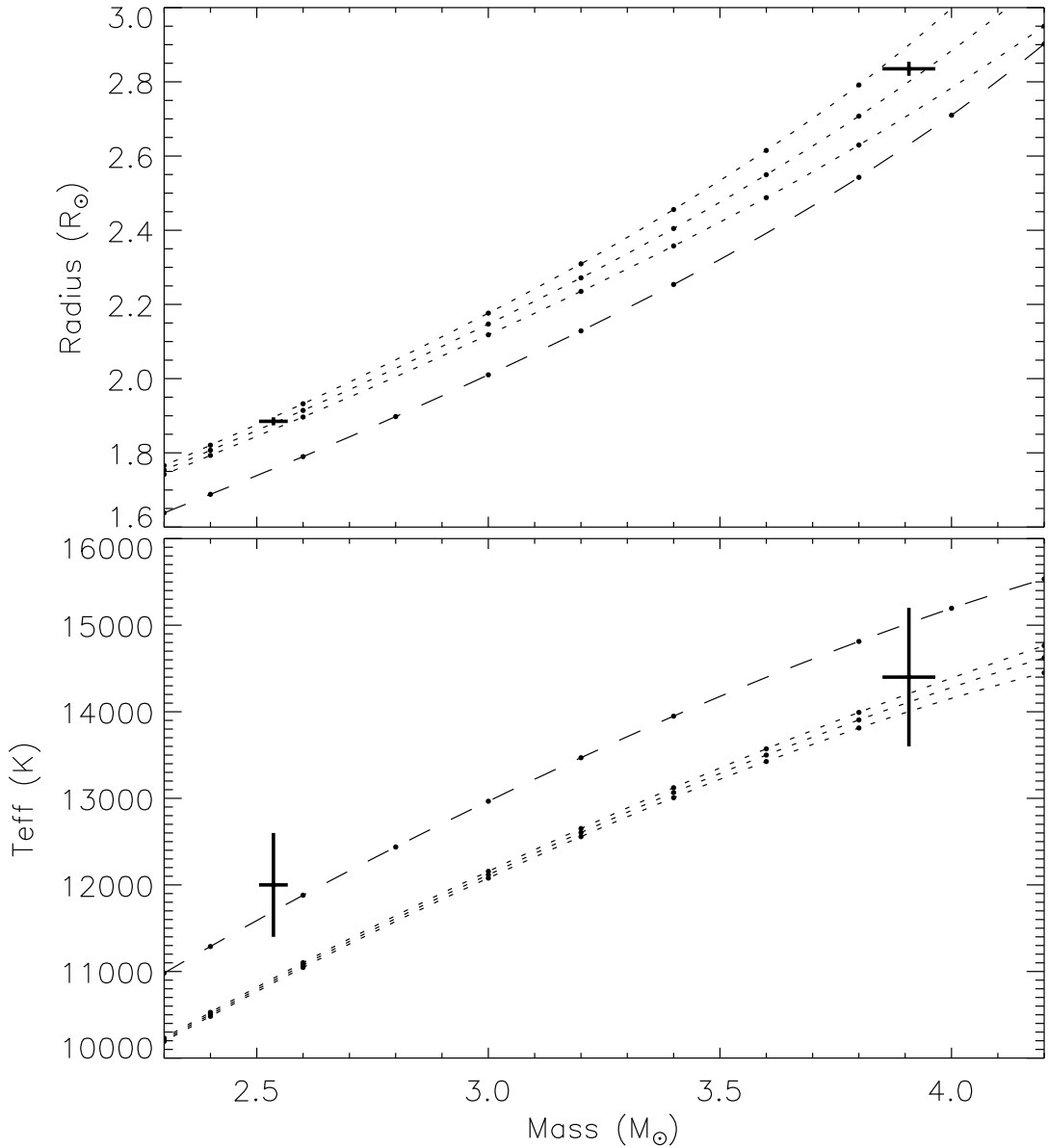} \\
\caption{\label{fig:theo} Mass--radius and mass--\Teff\ plots showing the properties of
$\zeta$\,Phe versus PARSEC models \cite{Bressan+12mn}. The dotted lines are
model predictions for $Z=0.017$ and ages of 70, 80 and 90 Myr. The dashed lines are
predictions for $Z=0.010$ and an age of 80\,Myr. The small circles show the points
at which the models have been tabulated; the lines have been generated by interpolation
between these point.} \end{figure}

The measured masses, radii and \Teff\ values have been compared to the predictions of several sets of theoretical models to gauge the level of agreement between observation and theory. The comparisons were performed in the mass--radius and mass--\Teff\ planes in order to have as direct a link as possible with the observational results \cite{MeClausen07aa}. An approximately solar metal abundance was assumed for the stars in the absence of direct spectroscopic measurements of their photospheric compositions.

The PARSEC models \cite{Bressan+12mn} fit the stars within the errorbars in the mass--radius diagram for a fixed fractional metal abundance of $Z = 0.017$ and an age of 80--90\,Myr. In the mass--\Teff\ plot the agreement is good for the primary star (0.5$\sigma$ lower than the measured value) but not for the secondary star (2$\sigma$ lower). A more metal-poor chemical composition improves the fit to the \Teff\ values but at the expense of the quality of fit in the mass--radius plane (see Fig.\,\ref{fig:theo}).

The Teramo models \cite{Pietrinferni+04apj} tell a similar story but using $Z=0.0198$ and giving an age of 70--80\,Myr. No significant difference to the fit is seen for models with and without convective core overshooting. Finally, the Yonsei-Yale models \cite{Demarque+04apjs} provide an essentially identical fit to the other grids of models.

$\zeta$\,Phe is therefore a young system with an approximately solar chemical composition, but the sets of theoretical models considered agree much better with each other than with the observed properties of this binary system. A detailed spectroscopic analysis of the two stars to determine precise \Teff\ values and photospheric chemical abundances would be useful.


\section*{Conclusion}

The bright eclipsing binary system $\zeta$\,Phoenicis has been analysed using the light curve of this system from the TESS mission and published RVs. The TESS light curve has essentially negligible Poisson noise and a perfect fit could not be obtained using the Wilson-Devinney code. The radii of the stars are nevertheless determined to precisions of better than 1\%, where the uncertainties were obtained by considering the best-fitting parameters found using a variety of plausible assumptions in the modelling process. The masses of the stars are measured to 1.5\% precision, using RVs measured from photographic spectra. The distance to the system determined from stellar radii, apparent magnitude and bolometric corrections agrees with that found using the \textit{Hipparcos} satellite. The masses, radii and \Teff\ of the primary star are well matched by theoretical predictions for a roughly solar metallicity and an age of 70--90\,Myr, but the \Teff\ of the secondary star is higher than predicted by approximately 2$\sigma$.

A new spectroscopic study of this system would be extremely helpful in refining the measurements of the masses and \Teff\ values of the stars, and establishing their photospheric chemical compositions and rotational velocities. Such spectra already exist in the archive of the European Southern Observatory (ESO) and are available for use. Once these results have been obtained, and because the two component stars have significantly different physical properties, $\zeta$\,Phe will be able to provide an exacting test of the predictions of stellar evolutionary theory.


\section*{Acknowledgements}

I would like to thank Pierre Maxted, Kresimir Pavlovski and an anonymous referee for their comments on this work at the draft stage. The following resources were used in the course of this work: the ESO archive; the NASA Astrophysics Data System; the SIMBAD database operated at CDS, Strasbourg, France; and the ar$\chi$iv scientific paper preprint service operated by Cornell University.


\section*{Afterword}

The writing of this paper began, partly by coincidence, on the day that I learned of the death of Johannes Andersen (1943--2020). Johannes was an inspiration in his careful analysis of so many eclipsing systems, and in his seminal 1991 review paper. I would like to dedicate this work to both Johannes, whom I knew only briefly, and to his colleague Jens Viggo Clausen (1946--2011). Jens Viggo was a wonderful and hospitable boss for my first postdoctoral position, at the Niels Bohr Institute in Copenhagen where both he and Johannes worked. \\


%
%
%
%
%
%
%
%


\end{document}